\overfullrule=0pt
\input harvmac

\def\s{{\sigma}}

\def\O{{\Omega}}

\def\half{{1\over 2}}
\def\p{{\partial}}

\Title{\vbox{\hbox{IFT-P.006/2000 }}}
{\vbox{
\centerline{\bf The Tachyon Potential in Open}
\centerline{\bf Neveu-Schwarz String Field Theory}}}
\bigskip\centerline{Nathan Berkovits\foot{e-mail: nberkovi@ift.unesp.br}}
\bigskip
\centerline{\it Instituto de F\'\i sica Te\'orica, Universidade Estadual
Paulista}
\centerline{\it Rua Pamplona 145, 01405-900, S\~ao Paulo, SP, Brasil}

\vskip .3in
A classical action for open superstring field theory has been proposed
which does not suffer from contact term problems. After generalizing
this action to include the non-GSO projected states of the Neveu-Schwarz
string, the pure tachyon contribution to the tachyon 
potential is explicitly computed. 
The potential has a minimum of $V = -{1\over{32 g^2}}$ which is $60\%$ 
of the predicted exact minimum of $V=-{1\over{2\pi^2 g^2}}$ from
D-brane arguments.

\Date {January 2000}

\newsec{Introduction}

The Neveu-Schwarz (NS) sector of the non-GSO projected superstring has recently
been reconsidered as part of a sensible physical theory \ref\sen
{A. Sen, {\it Stable Non-BPS States in String Theory} JHEP 9806 (1998) 007,
hep-th/9803194\semi
A. Sen, {\it Tachyon Condensation on the Brane Antibrane System} JHEP 9808 
(1998) 012,
hep-th/9805170.}\ref\susyb{
O. Bergman and M.R. Gaberdiel, {\it A Non-Supersymmetric Open String
Theory and S-Duality}, 
Nucl. Phys. B499 (1997) 183, hep-th/9701137\semi
T. Yoneya, {\it Spontaneously Broken
Space-Time Supersymmetry in Open String Theory without GSO Projection},
hep-th/9912255.}.
Although this sector contains a tachyon, there have been proposals
for removing the undesired properties of the the tachyon by
assuming a tachyon potential which is bounded from below.

The most efficient method for computing the tachyon potential uses
open string field theory\ref\kos{
V.A. Kostelecky and S. Samuel, {\it The Static Tachyon Potential in
the Open Bosonic String Theory}, Phys. Lett. B207 (1988) 169.}
\ref\sss{
A. Sen, {\it Universality of the Tachyon
Potential}, hep-th/9911116.}
\ref\szw{A. Sen and B. Zwiebach, {\it 
Tachyon Condensation in String Field Theory}, hep-th/9912249.},
however the cubic action of \ref\witsu{E. Witten, {\it Interacting
Field Theory of Open Superstrings}, Nucl. Phys. B276 (1986) 291.}
for open superstring field theory contains contact term problems
which spoil gauge invariance\ref\cont{C. Wendt, {\it Scattering
Amplitudes and Contact Interactions in Witten's Superstring Field Theory},
Nucl. Phys. B314 (1989) 209\semi J. Greensite and F.R. Klinkhamer,
{\it Superstring Amplitudes and
Contact Interactions}, Nucl. Phys. B304 (1988) 108.}. Recently, a new
action for open superstring field theory has been constructed \ref\new
{N. Berkovits, {\it Super-Poincare Invariant Superstring Field Theory},
Nucl. Phys. B450 (1995) 90, hep-th 9503099\semi N. Berkovits,
{\it A New Approach to Superstring Field Theory},
proceedings to the
$32^{nd}$ International Symposium Ahrenshoop on the
Theory of Elementary Particles, Fortschritte der Physik (Progress of
Physics) 48 (2000) 31, hep-th/9912121\semi
N. Berkovits and C.T. Echevarria, {\it
Four-Point Amplitude from Open Superstring
Field Theory}, hep-th/9912120.}
which does not
suffer from contact term problems. This action resembles a Wess-Zumino-Witten
action and can be naturally obtained by embedding the N=1 description of
the superstring into an N=2 string \ref\top{N. Berkovits and C. Vafa,
{\it N=4 Topological Strings}, Nucl. Phys. B433 (1995) 123, hep-th/9407190.}.

In this paper, the pure tachyon contribution to the tachyon potential will
be explicitly computed using this new action. The pure tachyon contribution
is 
\eqn\result{V(T) = -{1\over {4 g^2}} T^2 + {1\over {2 g^2}} T^4,}
which has a minimum of $V(T_0)=-{1\over {32g^2}}$ when $T_0=\pm \half.$
This value of the minimum is $60\%$ of the predicted
exact minimum of $V(T_0)=-{1\over{2\pi^2 g^2}}$ using D-brane arguments\foot
{In the original version of this paper, the mass of the brane-antibrane
was incorrectly stated to be ${1\over{\pi^2 g^2}}$. This value of the mass
is only correct if one doubles the number of states in the string field theory
action to allow for strings which end on the brane or antibrane \ref\pss
{A. Sen, private communication.} .}
where the mass of the brane-antibrane is ${1\over{2\pi^2 g^2}}$ \sss\pss.
It would be interesting to check if the remaining $40\%$ comes from
including contributions to
the effective tachyon potential from
non-tachyon fields, as was found for the bosonic string
tachyon potential in \szw.

\newsec{Neveu-Schwarz String Field Theory Action}

Using the superstring field theory action of \new,
the GSO-projected NS contribution is given by 
\eqn\nsc{S = 
{1\over{2 g^2}} Tr \langle (e^{-\Phi}Q e^{\Phi})(e^{-\Phi}\eta_0 e^{\Phi})
-\int_0^1 dt(e^{-t\Phi}\partial_t e^{t\Phi})\left\{e^{-t\Phi}Q e^{t\Phi},
e^{-t\Phi}\eta_0 e^{t\Phi}\right\} \rangle}
where $\eta_0 = \oint dz \eta(z)$ is defined by fermionizing the
super-reparameterization ghosts as\ref\fms
{D. Friedan, E. Martinec, and S. Shenker, {\it Conformal Invariance,
Supersymmetry, and String Theory}, Nucl. Phys. B271 (1986) 93.}
$\gamma=\eta e^\phi$ and $\beta=\p\xi e^{-\phi}$,  
$$Q=\oint dz [c ( T_{matter} - \eta\p\xi 
-\half\p\phi\p\phi - \p^2\phi -b\p c ) 
+\eta e^\phi G_{matter} -\eta\p\eta e^{2\phi}b],$$
and $\langle ~~\rangle$ signifies the two-dimensional correlation function
in the ``large'' RNS Hilbert space \fms\ where
$\langle \xi c \p c \p^2 c e^{-2\phi} \rangle =2$. 
The normalization of \nsc\ has been fixed by 
requiring that
the quadratic Yang-Mills contribution to the action is
$S = -{1\over {4 g^2}} Tr
\int d^{10} x  F_{mn} F^{mn}$, which is the correct sign for
the $(-+...+)$ metric that is being used.
String fields are multiplied using the midpoint interaction of \ref\witone
{E. Witten, {\it Noncommutative Geometry and String Field Theory}, Nucl. Phys.
B268 (1986) 253.} and $\Phi$ is related to the NS string field $V$ of 
\witsu\ by $\Phi= \xi_0 V$ or $V = \eta_0\Phi$.

In the GSO-projected sector, the NS string field $\Phi$ is bosonic.
Since the unprojected NS states are fermionic with respect
to the projected NS states, it will be convenient to define
$\widehat\Phi = \Phi_B \times I + \Phi_F \times \s_1$ where $\Phi_B$
described the projected states, $\Phi_F$ describes the unprojected states,
$I$ is the $2\times 2$ identity matrix, and $(\s_1,\s_2,\s_3)$
are the Pauli matrices \sss.
Furthermore, it will be convenient to define 
$$\widehat Q \equiv Q\times \s_3,\quad
\widehat \eta_0 \equiv \eta_0\times \s_3,$$
which satisfy $\widehat Q (\widehat\Phi_1 \widehat\Phi_2) = 
(\widehat Q \widehat\Phi_1)  
\widehat\Phi_2 + \widehat\Phi_1 (\widehat Q\widehat\Phi_2)$ and
$\widehat \eta_0 (\widehat\Phi_1 \widehat\Phi_2) = 
(\widehat\eta_0 \widehat\Phi_1) 
\widehat\Phi_2 + \widehat\Phi_1 (\widehat \eta_0\widehat\Phi_2)$.

The complete non-GSO projected NS string field theory action is defined by
\eqn\nongso{S = 
{1\over {4g^2}} Tr \langle (e^{-\widehat\Phi}\widehat Q e^{\widehat\Phi})
(e^{-\widehat\Phi}\widehat\eta_0 e^{\widehat\Phi})
-\int_0^1 dt(e^{-t\widehat\Phi}\partial_t e^{t\widehat\Phi})
\left\{e^{-t\widehat\Phi}\widehat Q e^{t\widehat\Phi},
e^{-t\widehat\Phi}\widehat\eta_0 e^{t\widehat\Phi}\right\} \rangle}
where the trace is over the $2\times 2$ matrices as well as the
Chan-Paton matrices.

One can check that \nongso\ is invariant under
the WZW-like gauge transformation
\eqn\gauge{\delta e^{\widehat\Phi} = (\widehat Q\widehat\Omega)
 e^{\widehat\Phi} + e^{\widehat\Phi} (\widehat\eta_0\widehat\Omega')}
where $\widehat\Omega$ and $\widehat\Omega'$ are string fields
of the form $\widehat\Omega=\O_F \times \s_3 +i\Omega_B\times \s_2$ 
with $\O_F$ being fermionic
and projected while $\O_B$ is bosonic and unprojected.
One subtle point in proving this gauge invariance is that 
$\langle \widehat\Phi_1\widehat\Phi_2\rangle=
\langle \widehat\Phi_2\widehat\Phi_1\rangle $ since 
when $A$ and $B$ are unprojected states,
$\langle A B \rangle= \mp \langle B A\rangle$
where the minus sign is if they are bosons and the plus sign is if they
are fermions. This reversal of the usual statistics comes from square-root
factors produced by the $\half$-integer conformal weight of unprojected
NS states. Note that a similar subtlety occurs with unprojected states
using the action of \witsu.

\newsec{Computation of Tachyon Potential}

Expanding the action of \nongso\ in powers of $\widehat\Phi$, one obtains
\eqn\expand{ S  ={1\over{2 g^2}} Tr \langle \half (\widehat Q \widehat\Phi)   
(\widehat \eta_0 \widehat\Phi)   
-{1\over 6} 
\widehat\Phi\{\widehat Q\widehat\Phi~,~\widehat\eta_0\widehat\Phi\} 
-{1\over{24}}
[\widehat\Phi,\widehat Q\widehat\Phi][\widehat\Phi,\widehat\eta_0
\widehat\Phi] + ... \rangle.}
To compute the term with $N$ $\widehat\Phi$'s, one uses the map 
$$w(z) = \left({{1-iz}\over {1+iz}}\right)^{2\over N}$$
from the disc to a $2\pi/N$ wedge of the complex plane.
Rotating this map by a factor $e^{{{2\pi i}\over N}}$ 
allows each successive string
field to get mapped to a different $2\pi/N$ wedge. The center
of the $J^{th}$ disc gets mapped to the point $e^{{{2\pi i (J-1)}\over N}}$ and,
to obtain an SL(2,R)-invariant expression,  the $J^{th}$
string field gets multiplied
by a factor $(e^{{{2\pi i (J-1)}\over N}} {4\over{Ni}})^h$
where $h$ is the conformal weight of the string field
and ${4\over{Ni}}$ is
${{dw}\over{dz}}|_{z=0}$
\ref\sft{
E. Cremmer, A. Schwimmer and C. Thorn, {\it The Vertex Function in
Witten's Formulation of String Field Theory}, Phys. Lett. B179 (1986) 57\semi
D.J. Gross and
A. Jevicki, {\it Operator Formulation of Interacting
String Field Theory}, Nucl. Phys. B283 (1987) 1\semi
A. LeClair, M.E. Peskin and
C.R. Preitschopf, {\it String Field Theory on the Conformal Plane (I)},
Nucl. Phys. B317 (1989) 411.}.

The tachyon field $T(x)$ appears in the string field $\widehat\Phi$ as
\eqn\app{\widehat\Phi = i\xi c e^{-\phi} T(x) \times \s_1}
where the factor of $i$ is needed to get the right sign for the kinetic
term.
One can easily compute that at zero momentum, 
\eqn\qphi{\widehat Q \widehat\Phi =
  T (\half c\p c \xi e^{-\phi} +\eta e^\phi)
\times \s_2,\quad
\widehat \eta_0 \widehat\Phi = -  T c e^{-\phi}  \times \s_2.}

Since $\langle \xi c \p c \p^2 c e^{-2\phi}  \rangle =2$, the only
pure tachyon contribution to the action of \nongso\ comes from the
quadratic and quartic terms of \expand. The quadratic contribution
to the action (which is minus the tachyon potential) is given by
\eqn\quadr{ - V_2 = -
{1\over {2 g^2}}
T^2 \langle (\half c\p c \xi e^{-\phi}(1))(c e^{-\phi}(-1))\rangle
(-2i)^{-\half}
(2i)^{-\half} = {1\over {4g^2}} T^2.}

The quartic contribution to the action is given by 
\eqn\quartic{- V_4= {{T^4}\over{24 g^2}}
(-i)^{-\half}
(1)^{-\half}
(i)^{-\half}
(-1)^{-\half} 
\langle(\xi c e^{-\phi}(1)) (\eta e^\phi(i)) (\xi c e^{-\phi}(-1))
(c e^{-\phi}(-i))  }
$$
+ (\eta e^\phi(1))(\xi c e^{-\phi}(i)) (\xi c e^{-\phi}(-1)) (c e^{-\phi}(-i)) 
+(\xi c e^{-\phi}(1)) (\eta e^\phi(i)) 
(c e^{-\phi}(-1))(\xi c e^{-\phi}(-i)) $$
$$
+
(\eta e^\phi(1))(\xi c e^{-\phi}(i)) 
(c e^{-\phi}(-1))(\xi c e^{-\phi}(-i)) \rangle = -{{T^4}\over {2 g^2}}.$$

So $V(T) = V_2 + V_4 = -{1\over {4 g^2}}T^2 +{1\over {2 g^2}} T^4$ 
which has a minimum of 
$V(T_0)=-{1\over {32g^2}}$ when $T_0=\pm\half $.

{\bf Acknowledgements:} I would like to thank Oren Bergman,
Ashoke Sen, Ion Vancea
and Barton Zwiebach for useful discussions, Caltech for
their hospitality, and CNPq grant 300256/94-9
for partial financial support.

\listrefs

\end